\documentclass[superscriptaddress,reprint,aps,prl]{revtex4-2}
\usepackage[usenames,dvipsnames,table]{xcolor}
\usepackage[utf8]{inputenc}
\usepackage{graphicx}
\usepackage{bbold}
\usepackage{hyperref}
\usepackage{array}
\usepackage{setspace}
\usepackage{amsthm}
\usepackage{amsfonts}
\usepackage{amsmath}
\usepackage{bbm}
\usepackage{colonequals}
\usepackage{mathtools}
\usepackage{tikz}
\usepackage{xspace}
\usepackage{braket}
\usepackage[normalem]{ulem}

\newtheorem*{result*}{Result}

\newtheorem*{conj*}{Conjecture}

\def\zb{{\bar{z}}}
\def\wb{{\bar{w}}}

\def\GG{{\mathcal{G}}}

\def\NN{{\mathcal{N}}}
\def\OO{{\mathcal{O}}}

 \let\h=\eta   
    
   \let\f=\phi  
  \let\D=\Delta

\newcommand{\be}{\begin{equation}}
\newcommand{\ee}{\end{equation}}
\newcommand{\bea}{\begin{eqnarray}}
\newcommand{\eea}{\end{eqnarray}}

\let\<=\langle
\let\>=\rangle
\newcommand{\vev}[1]{\ensuremath{\langle #1 \rangle}\xspace}

\begin{document}

\title{QFT in AdS instead of LSZ}

\author{Balt C. van Rees}
\affiliation{CPHT, CNRS, \'Ecole Polytechnique, Institut Polytechnique de Paris, Route de Saclay, 91128 Palaiseau, France}

\author{Xiang Zhao}
\affiliation{CPHT, CNRS, \'Ecole Polytechnique, Institut Polytechnique de Paris, Route de Saclay, 91128 Palaiseau, France}
\affiliation{Fields and String Laboratory, Institute of Physics, Ecole Polytechnique F\'ed\'erale de Lausanne (EPFL), Rte de la Sorge, CH-1015 Lausanne, Switzerland}

\begin{abstract}
The boundary correlation functions for a QFT in a fixed AdS background should reduce to S-matrix elements in the flat-space limit. We consider this procedure in detail for four-point functions. With minimal assumptions we rigorously show that the resulting S-matrix element obeys a dispersion relation, the non-linear unitarity conditions, and the Froissart-Martin bound. QFT in AdS thus provides an alternative route to fundamental QFT results that normally rely on the LSZ axioms.
\end{abstract}

\maketitle

\section{Introduction}
Consider a gapped QFT in AdS$_{d+1}$ with curvature radius $R$. As discussed in detail in \cite{Paulos:2016fap}, such a setup defines a set of conformally invariant boundary correlation functions which for any $R$ obey all the usual CFT axioms except for the existence of a stress tensor. Furthermore, as we take the flat-space limit $R \to \infty$ these boundary correlation functions should transmogrify in some sense into the S-matrix of the QFT. This idea dates back to \cite{Polchinski:1999ry} and the massive case has subsequently been explored in \cite{Giddings:1999qu,TTBar,Paulos:2016fap,Carmi:2018qzm,Hijano:2019qmi,Komatsu:2020sag,Li:2021snj,Gadde:2022ghy}. The various prescriptions which have emerged have been checked in numerous examples, including some non-perturbative matches between S-matrix and conformal bootstrap results \cite{Paulos:2016fap,Paulos:2016but,Homrich:2019cbt}.

The QFT in AdS approach is remarkable because it offers a new \emph{axiomatic} way to obtain S-matrix elements as limits of mathematically well-defined conformal correlation functions. It is in particular entirely orthogonal to the textbook LSZ prescriptions. As reviewed for example in \cite{Bogolyubov:1990kw}, the corresponding \emph{LSZ axioms} lie at the basis of all the foundational results in S-matrix theory, including the known domains of analyticity of scattering amplitudes as well as physical constraints like the Froissart-Martin bound. It is worthwhile to investigate whether the QFT in AdS approach can recover or even extend these foundational results, not in the least to strengthen our belief in the LSZ axioms and the mathematical consistency of QFT in curved and flat space.

In this paper we take a first step in this direction for four-point functions of identical boundary scalar operators corresponding to a light particle. After making one technical assumption, namely that the flat-space limit is finite in a subset of the Euclidean domain, we show that such a four-point function always produces a consistent scattering amplitude in the sense that it obeys all the known physical constraints of unitarity, analyticity, and crossing symmetry.

Our results on analyticity have been made possible by the remarkable progress in our understanding of CFT correlation functions, starting with the Lorentzian inversion formula \cite{Caron-Huot:2017vep}. We will find good use for the Polyakov-Regge blocks \cite{Mazac:2019shk,Gopakumar:2016wkt,Gopakumar:2018xqi,Dey:2016mcs,Dey:2017fab,Dey:2017oim}, in particular the subtracted version presented in \cite{Caron-Huot:2020adz} that can be obtained from the conformal dispersion relation of \cite{Carmi:2019cub}. We briefly discuss the flat-space limit of these formulas at the end of this paper, leaving a more detailed presentation to an upcoming companion paper \cite{xiangpaper}.

We will always assume $d > 1$. See \cite{Cordova:2022pbl} for a similar approach to the $d=1$ case.

\section{Conformal Mandelstam variables}
We consider a \emph{family} of conformally covariant four-point functions of boundary scalar operators $\phi$ with dimension $\Delta_\phi$, labelled by a parameter $R > 0$. We consider a gapped bulk theory, and therefore the \emph{flat-space limit} $R \to \infty$ always implies that all scaling dimensions $\Delta_i \to \infty$. We will assume that $\phi$ generates a single-particle state such that the ratio $\lim_{R \to \infty} \Delta_\phi /R = m$ remains finite. (In the following we will set $m = 1$.) We will also assume that the particle is \emph{light}, in the sense that the next primary operator has a scaling dimension
\begin{equation}
	\Delta' > \sqrt{2} \Delta_\phi
	\label{spectrum assumption}
\end{equation}
for sufficiently large $R$ \footnote{In the flat-space limit this spectrum assumption means that the scattering amplitude cannot develop anomalous thresholds. Extending the results of this work to more general cases is an open question.}.

Our kinematical conventions are as follows. The four-point function takes the form:
\begin{equation}
	\vev{\phi_1 \phi_2 \phi_3 \phi_4} = x_{12}^{-2 \D_\phi} x_{34}^{-2\D_\phi} \GG(s,t,u)
\end{equation}
As indicated by the notation, we work with the \emph{conformal Mandelstam invariants} $(s,t,u)$, defined as \cite{Komatsu:2020sag}:
\begin{equation}
	r(s) = \frac{2 - \sqrt{s}}{2 + \sqrt{s}},\, \eta(s,t) = - 1 + \frac{2t}{4 -s}, \, u = 4 - t - s
	\label{eq: conformal Mandelstam}
\end{equation}
where $r = \sqrt{\rho \bar \rho}$, $\eta = (\rho + \bar \rho)/ (2 \sqrt{\rho \bar \rho})$ and $\rho$, $\bar \rho$ are the radial coordinates of \cite{Hogervorst:2013sma}. 
For more details see \footnote{See Supplemental Material for an exposition of these variables.}.


We also introduce an involution operation which we denote with a tilde:
\begin{equation}
	(\tilde s, \tilde t, \tilde u) \colonequals \left(16, - 4u,- 4 t\right)/s
	\label{eq: involution}
\end{equation}
such that $r(\tilde s) = - r(s)$ and $\eta(\tilde s,\tilde t) = - \eta(s,t)\equalscolon \tilde{\h}$. This operation is a bijection between the original Euclidean region and the $s$-channel physical region. 

\section{The flat-space limit}
To extract a scattering amplitude from the correlation function we introduce the \emph{pre-amplitude} $T(s,t,u)$ via the decomposition:
\begin{equation}
\label{Tdefn}
	 \GG(s,t,u) = \GG_\text{gff}(s,t,u) + \GG_c(s,t,u) T(s,t,u)
\end{equation}
with the disconnected correlator:
\begin{align}
	&\quad\GG_\text{gff}(s,t,u) =\\ &(4-s)^{2\D_\phi} \left((4-s)^{-2\D_\phi} + (4-t)^{-2\D_\phi} + (4-u)^{-2\D_\phi} \right) \nonumber
\end{align}
and the large $R$ limit of a (rescaled) contact diagram:
\begin{equation}
	\GG_c(s,t,u) = \frac{w_c 2^{4 - 4 \Delta_\phi}  (4-s)^{2\D_\phi}}{\sqrt{(4-s)(4-t)(4-u)}}
	\label{flat-space limit of contact}
\end{equation}
with a normalization $w_c = 2^{-\frac{1}{2} (d+7)} \pi ^{\frac{1-d}{2}} \Delta_\phi ^{\frac{d-5}{2}} R^{3-d}$. The expectation is now more or less that
\begin{equation}
\label{Tfiniteassumption}
\lim_{R \to \infty}T(s,t,u) = T_\infty (s,t,u)
\end{equation}
produces a \emph{bona fide} scattering amplitude. More precisely, there are several heuristic derivations showing that the above claim holds some truth \cite{TTBar,Hijano:2019qmi,Komatsu:2020sag}. When tested on Witten diagrams it was however shown in \cite{Komatsu:2020sag} that \emph{AdS Landau diagrams} produce regions in the Mandelstam plane where the large $R$ limit diverges instead.

In the flat-space limit exponential differences arise from all the $(\cdot)^{\Delta_\phi}$ terms in the above expressions. Note however also the scaling $w_c \sim R^{\frac{1-d}{2}}$, which for $d > 1$ implies that the connected correlator must have a further power-law suppression to produce a finite amplitude. This is an AdS avatar of the flat-space fact that only in $d = 1$ the connected and disconnected part of a two-to-two S-matrix element are equally singular distributions. Our main assumption will now be that this power-law suppression of the connected correlator holds at least in a subregion of the Euclidean domain. More precisely, if
\begin{equation}
\label{E prime region}
	E' = \{(s,t,u) \,|\, s,t,u \leq 2  \text{ and } s + t + u = 4  \}\,,
\end{equation}
then \emph{we will assume that $T_\infty$ as defined in \eqref{Tfiniteassumption} exists pointwise for all $(s,t,u) \in E'$}. We will show that with this assumption in place we can construct a consistent unitary scattering amplitude with a large domain of analyticity.

\section{Conformal dispersion relation}
Following the logic leading to the Lorentzian inversion formula, the authors of \cite{Carmi:2019cub} wrote down a \emph{conformal dispersion relation}. In its standard form it expresses the Euclidean correlator as an integral transform of the double discontinuity of the Lorentzian correlator. For physical correlation functions this dispersion relation does not always converge, but various subtraction procedures exist to compensate for this issue. We will use the subtracted dispersion relation described in \cite{Caron-Huot:2020adz}, which was inspired by a Mellin space \cite{Mack:2009mi,joaomellin} analysis.

The (subtracted) conformal dispersion relation reads:
\begin{multline}
\label{confdispreln}
	(z \zb)^{-\Delta_\phi} \left[\GG(z,\zb) - \GG_\text{gff}(z,\zb)\right] =\\
	 \iint dw d\wb \, K_2(z,\zb;w,\wb) \text{dDisc}_s[(w \wb)^{-\Delta_\phi} (\GG(w,\wb) -1)] \\+ \left((z,\zb) \leftrightarrow (1-z, 1-\zb)\right)
\end{multline}
where the exact expression for the kernel $K_2(z,\zb;w,\wb)$ and the integration region can be found in \cite{Caron-Huot:2020adz}. The dDisc$_s[\cdot]$ operator corresponds to taking the $s$-channel double discontinuity of a conformal correlation function, an operation first formalized in \cite{Caron-Huot:2017vep} (and which should not be confused with the Mandelstam double discontinuity of a scattering amplitude). The subtraction implicit in the kernel ensures that this integral is always finite in the 
Regge limit. A potential divergence can however still arise from the so-called lightcone limit on the second sheet; 
in the following we assume this to be under control for all finite $R$ \footnote{This assumption is implicit in much of the literature on the analytic conformal bootstrap (see for example \cite{,Hartman:2015lfa,Caron-Huot:2017vep,Simmons-Duffin:2017nub}). Numerical evidence for 3d Ising \cite{Caron-Huot:2020ouj} and the (2,0) theories \cite{Lemos:2021azv} indicates that there are no obvious contradictions if the lightcone limit on the second sheet is assumed to be regular. If one were to assume only the rigorously proven bound of conformal correlators in the lightcone limit on the second sheet then our arguments would require a conformal dispersion kernel with more subtractions, leading to a dispersion relation similar to \eqref{positivedispersionrelation} with more subtractions.}. Finally we note that the above representation is singular as $w \to 0$ if there are conformal blocks with $\Delta - \ell < 2 \Delta_\phi$, but this can be mitigated by a small deformation of the integration contour.

The contribution of a single conformal block $G_\Delta^{(\ell)}(u,v)$ to the conformal dispersion relation is known as a Polyakov-Regge block. If we introduce:
\begin{multline}
	\label{blockcontrtoG}
\iint dw d\wb \, K_2(z,\zb;w,\wb) \text{dDisc}_s[(w \wb)^{-\Delta_\phi} G_\Delta^{(\ell)}(w,\wb)]
	=\\  (z \zb)^{-\Delta_\phi} G_c(z,\zb) T_\Delta^{(\ell)}(z,\zb)
\end{multline}
then we can use the swappability property of the kernel for fixed $(z,\zb)$ (in the sense of \cite{Qiao:2017lkv}) to write, now using the conformal Mandelstam variables: 
\begin{equation}
	\label{PRblockdecomposition}
	T(s,t,u) = \sum_{\OO \neq 1} a_\OO^2 T_{\Delta_\OO}^{(\ell_\OO)}(s,t,u) + (s \leftrightarrow t)
\end{equation}
where the sum is over all non-identity operators and $a_\OO^2$ is the coefficient of the corresponding conformal block. Like the original conformal block decomposition, this subtracted Polyakov-Regge block decomposition is expected to be absolutely convergent (at finite $R$) in a large domain that includes the entire Euclidean Mandelstam triangle. 


We are interested in the large $\Delta$ limit of the Polyakov-Regge blocks. As is discussed for example in \cite{Kos:2013tga}, the large $\Delta$ 
limit of an $s$-channel conformal block reads \footnote{For gapped QFTs in AdS both $h = (\Delta - \ell)/2$ and $\bar h = (\Delta + \ell)/2$ become large in the flat-space limit. In $d = 2$ and $d = 4$ one readily verifies that equation \eqref{largeDblock} is correct in the sense that the double limit $h,\bar h \to \infty$ of the ratio of the left- and the right-hand side equals 1 for each fixed $r$, with corrections that are suppressed by a simple function of $\min(h,\bar h)$. This more general statement is likely true for all $d$.}
\begin{equation}
\label{largeDblock}
	G_{\Delta}^{(\ell)}(s,t,u) \overset{\Delta \to \infty}{\sim} \frac{ \sqrt{s} \left(s^{1/4}+\tilde s^{1/4}\right)^d }{\sqrt{(4-t) (4-u)}} (4r(s))^{\Delta }  P_\ell^{(d)}(\eta)\,.
\end{equation}
The integrals for the Polyakov-Regge block can then be computed with a saddle-point computation, which yields:
\begin{widetext}
\begin{align}
	\frac{T_\Delta^{(\ell)}(s,u)}{(4 - s) (s + u)}  \overset{\Delta \to \infty}{\sim}  \frac{(\Delta_\phi/R)^{-2}}{\left(\mu ^2-4\right)} \times  \frac{1}{\Xi^{(\ell)}_{\Delta,\Delta_\phi}} \times \frac{P_\ell^{(d)}\left( -1 + \frac{2 u}{4 - \mu^2} \right)}{\left(s-\mu ^2\right) \left(\mu ^2 + u\right)} + \frac{E_\Delta^{(\ell)}(s,u)}{(4-s)(s+ u)}
\label{blockcontrtoT}
\end{align}
\end{widetext}
with $\mu \colonequals \Delta/\Delta_\phi$. Let us postpone discussing the `error' term $E_\Delta^{(\ell)}(s,u)$. The non-trivial scaling of the first term in the flat-space limit is then entirely absorbed in the factor $\Xi^{(\ell)}_{\Delta,\Delta_\phi}$, which is just the coefficient of the single-trace conformal block in an $s$-channel exchange Witten diagram with unit bulk coupling \footnote{To compute $\Xi^{(\ell)}_{\D, \D_\f}$ one can write down the conformal partial wave decomposition of the exchange Witten diagram using the spectral representation of the bulk-bulk propagator (see Appendix B of \cite{joaomellin} for scalar exchange and Section 6 of \cite{Costa:2014kfa} for spinning exchange). This density has a pole at the location of the single-trace block whose residue is exactly $\Xi^{(\ell)}_{\D, \D_\f}$}. 
The last factor of the first term is then just 
the flat-space $s$-channel exchange diagram with the given subtractions. We can therefore write \footnote{Equation \eqref{Tdispersion} is obtained by substituting the right-hand side of equation \eqref{blockcontrtoT} into equation \eqref{PRblockdecomposition}. We will assume the pointwise validity of this expression for $(s,u)$ inside $E'$. This would follow if the subleading corrections to (the ratio of the right- and the left-hand side of) \eqref{blockcontrtoT} are uniformly small for all $h$ and $\bar h$. We believe this can be proved rigorously, but it is not immediate: the large $\Delta$ limit of the conformal block as given in equation \eqref{largeDblock} is not uniform in $r$ and so we cannot automatically swap it with the integration kernel in the conformal dispersion relation.}
\begin{multline}
\label{Tdispersion}
	\frac{T^\text{sub}(s,t,u)}{(4-s)(s+u)} = \\
	\sum_\ell \int d\mu \, \frac{\rho_\ell(\mu)(2 \mu ^2+u-4) P_\ell^{(d)}\left( -1 + \frac{2 u}{4 - \mu^2} \right)}{(\mu^2 - 4) \left(\mu ^2 + u\right)\left(s-\mu ^2\right) \left(\mu ^2+s+u-4\right)}\\
	+ \text{subleading}
\end{multline}
where the `sub' superscript implies that we subtracted the error term, and with a positive spectral density $\rho_\ell(\mu)$ given by:
\begin{equation}
	\rho_\ell(\mu) = \sum_{\substack{\OO\text{ with spin }\ell\\\OO \neq 1}}  \frac{a_\OO^2}{m^2 \,\Xi^{(\ell)}_{\Delta,\Delta_\phi}} \delta(\mu - \Delta_\OO/\Delta_\phi) 
\end{equation}
which features in particular the conformal block coefficients $a_\OO^2$. If we consider  $4 - u - \mu^2 < s < \mu^2$ inside $E'$, then 
the integrand is not sign-definite: contributions for $\mu^2 > 4$ are always negative, whereas for $\mu^2 < 4$ the contributions are positive until the Gegenbauer polynomials start oscillating for $\mu^2 < 4 - u$. 
(The apparent singularity at $\mu^2 = 4$ is offset by a double zero at $\rho_\ell(\mu)$ that arises from the $1/\Xi^{(\ell)}_{\Delta,\Delta_\phi}$ factor.)

It is not immediately clear that the above sum-plus-integral remains convergent in the flat-space limit. However if we choose $(s_1,u)$ and $(s_2,u)$ inside $E'$ then for the expression
\begin{widetext}
\begin{equation}
\label{positivedispersionrelation}
	\frac{T^\text{sub}(s_1,u) - T^\text{sub}(s_2,u)}{(s_1-s_2)(s_1 + s_2 + u - 4)} = \sum_\ell \int d\mu \frac{\rho_\ell(\mu) (2 \mu ^2+u-4) P_\ell^{(d)}\left( -1 + \frac{2 u}{4 - \mu^2} \right)}{(s_1 - \mu^2)(s_2 - \mu^2)(4 - u - s_1 - \mu^2)(4-u-s_2 - \mu^2)}
\end{equation}
\end{widetext}
the integrand comes out to be non-negative as long as $4 - \mu^2 \leq u \leq 4$ for all $\mu$ for which the integral has support; by our assumption \eqref{spectrum assumption} this in particular includes the maximal value $u = 2$ that is still inside $E'$. Since the left-hand side remains finite in the flat-space limit by assumption, the right-hand side cannot diverge, either \footnote{From this argument it follows that demanding finiteness in all of $E'$ is actually not necessary. Instead we only need finiteness at two suitably chosen points $(s_1,u)$ and $(s_2,u)$, or (by sending $s_2 \to s_1$) even just finiteness of the derivative $\partial_s T^{\text{sub}}(s,u)$ at a single point. Such a more minimalist assumption might be useful in numerical studies of the flat-space limit.}. But this non-divergence implies that the limit function $T^\text{sub}_\infty(s_1,u)$ is actually \emph{analytic} everywhere in the complex $s$-plane with the exception of the $s$- and $t$-channel cuts starting at $\mu_0^2$ and $4 - u - \mu_0^2$ \footnote{To see this, map the cut $s$-plane to the unit disk with a coordinate $\chi$. The positivity of the density $\rho_\ell(\mu)$ implies that the coefficients of the series expansion at $\chi = 0$ are bounded in terms of the value of the function at $\chi = 0$, so within the disk we find that the function is essentially bounded in modulus by a multiple of $1/(1- |\chi|)$. This bound is uniform on compact subsets, so the limiting function is analytic by Montel's theorem. The power-law bound near the edge of the disk further implies the existence of a fixed-$u$ dispersion relation, although possibly with one more subtraction than the function at finite $R$. Note that we do not (need to) claim that the flat-space limit can be swapped with the sum-plus-integral.}. 
An identical result now follows for all $u\in[0,2]$ since the Gegenbauer polynomials $P_\ell^{(d)}(z) \leq P_\ell^{(d)}(1)$ for all $- 1 \leq z \leq 1$. Thus $T^\text{sub}_\infty(s,u)$ is our candidate analytic scattering amplitude.

Let us now discuss the error term $E_\Delta^{(\ell)}(s,t,u)$. It arises in exactly the same way as the AdS Landau diagram contributions to the Witten exchange diagram discussed in \cite{Komatsu:2020sag}, and its flat-space limit is either zero or infinite. In fact, when it diverges it does so in precisely the same manner as a conformal block itself, so:
\begin{equation}
	E_\Delta^{(\ell)}(s,t,u) = 
	\begin{cases}
	G_\Delta^{(\ell)}(s,t,u)/\GG_c(s,t,u) & s \in D_\mu\\
	0 & s \notin D_\mu
	\end{cases}
\end{equation}
where the problematic region $D_\mu$ is a compact ellipsoidal region contained in the disk $|s - 4| \leq 4 - \mu^2 $; see for example figure 17 in \cite{Komatsu:2020sag} for an illustration. In particular, $D_\mu$ is empty for $\mu > 2$ so \emph{only conformal blocks below the two-particle threshold produce an error term}.

The original $T_\infty(s,u)$ includes these error terms, and therefore diverges in the union of all the $D_\mu$ regions as well as its images under crossing. By simply throwing away these error terms we obtained the function $T_\infty^\text{sub}(s,u)$ which we have shown remains finite and analytic. Note that we only subtracted $s$- and $t$-channel error terms, but the $u$-channel error terms do not matter for $u\in E'$ due to our spectrum assumption \eqref{spectrum assumption}.

Let us finally note that the positivity of the spectral density $\rho_\ell(\mu)$ is a property known as \emph{extended unitarity} which to the best of our knowledge had never been axiomatically proven, but was essential for S-matrix bootstrap studies like that of \cite{Homrich:2019cbt}.

\section{Hyperbolic partial waves}

The aim of this section is to show that the analytic amplitude that we found also obeys the non-linear unitarity condition for $s > 4$.

We first define the hyperbolic partial waves (for the $s$-channel) as: 
\begin{equation}
	c_\ell(s) \coloneqq \frac{\NN_d}{2} \int_{-1}^1 d\eta (1-\eta^2)^{\frac{d-3}{2}} P_\ell^{(d)}(\eta) \frac{\GG(s,\eta) - 1}{\GG_c(s,\eta)}
\end{equation}
with $\NN_d = (16 \pi)^{-h}/2 \Gamma(h)$ with $h = (d-1)/2$. For any physical correlator $\GG(s,\eta)$ the hyperbolic partial waves are analytic functions in the complex $s$ plane minus the cuts starting at $s > 4$ and $s < 0$, which respectively correspond to $r < 0$ and $|r| = 1$. When we evaluate $c_\ell(s)$ for $s > 4$ it will be understood that we are slightly above the cut, which corresponds to physical kinematics.

For the disconnected parts of the correlator we use:
\begin{multline}
	\lim_{R \to \infty} \frac{\NN_d}{2} \int_{-1}^1 d\eta (1-\eta^2)^{\frac{d-3}{2}} \frac{P_\ell^{(d)}(\eta) }{\GG_c(s,\eta)} \left( \frac{1-s/4}{1-t/4} \right)^{2\Delta_\phi} = \\
	 - i (-1)^\ell \frac{1}{2} \sqrt{s} (s - 4)^{1 - d/2}\,,
\end{multline}
since the integral localizes near $t = 0$, so $\eta = -1$. By symmetry, if we exchange $t$ and $u$ then we will find the same expression without the $(-1)^\ell$. Altogether we can then write, for even $\ell$,
\begin{equation}
\label{hyppartialwavefromf}
	\lim_{R \to \infty} c_\ell(s) = - i \sqrt{s} (s - 4)^{1 - d/2} + f_\ell(s)
\end{equation}
where $f_\ell(s)$ are by definition the hyperbolic partial waves for the connected correlation function:
\begin{equation}
	f_\ell(s) \colonequals \frac{\NN_d}{2} \int_{-1}^1 d\eta (1-\eta^2)^{\frac{d-3}{2}} P_\ell^{(d)}(\eta) T(s,u(s,\eta))
\end{equation}

Below we will compare the hyperbolic partial waves against the reflected hyperbolic conformal partial waves:
\begin{equation}
\label{cell}
 	\tilde c_\ell(s) \coloneqq \frac{\NN_d}{2} \int_{-1}^1 d\eta (1-\eta^2)^{\frac{d-3}{2}} P_\ell^{(d)}(\eta) \frac{\GG(\tilde s,\tilde \eta) - 1}{\GG_c(s,\eta)}
\end{equation}
We will always evaluate these for physical $s > 4$ and $-1 \leq \eta \leq 1$. In that case $\tilde s$ lies in the Euclidean region and the numerator is real and free of branch cut ambiguities.

The relevance of the reflected hyperbolic conformal partial waves is as follows. We claim that, in the flat-space limit:
\begin{equation}
\label{cctildeinequality}
	|\tilde c_\ell(s)| \geq |c_\ell(s)|\,.
\end{equation}
This simply follows from the $s$-channel conformal block decomposition of both $c_\ell(s)$ and $\tilde c_\ell(s)$ and the large-$\Delta$ limit of the conformal block 
\eqref{largeDblock} and the contact diagram \eqref{flat-space limit of contact}. The integrals simply project onto the right spin, resulting in the following contributions of a conformal block to each of the hyperbolic partial waves:
\begin{equation}
	\tilde{c}_\ell(s) \supset \text{pref}(s) (-4r(s))^\Delta, \quad c_{\ell}(s) \supset \text{pref}(s) (4r(s))^\Delta\,,
\end{equation}
where $\text{pref}(s)$ is an unimportant positive prefactor \footnote{Notice that this prefactor is only manifestly positive after taking the limit $\D\to\infty$. It is not clear to us whether to which extent it remains positive for finite $\Delta$. One way to potentially prove it is to show that $(\GG(s,\h)-1)/\GG_c(s,\h)$ and its reflection in \eqref{cell} have a decomposition in Gegenbauer polynomials with positive coefficients.}. The region $s > 4$ corresponds to $-1 < r < 0$ and so the $s$-channel conformal block decomposition converges, leading immediately to equation \eqref{cctildeinequality}.

\subsection{The trimmed amplitude}
To discuss the emergence of unitarity (in the scattering amplitude sense) we introduce yet another function. Let us write:
\begin{multline}
\label{Tcutdefinition}
	\GG_\text{gff}(s,t,u) + \GG_c(s,t,u)T^{\text{trim}}(s,t,u) = \\
	 1 + \sum_{\Delta_{\OO} \geq 2 \Delta_\phi,\ell_\OO} a_\OO^2 G_{\Delta_\OO}^{(\ell_\OO)}(s,t,u)
\end{multline}
Trimming all the non-trivial conformal blocks below threshold in the $s$-channel is a rather dramatic operation that destroys for example crossing symmetry, Regge-boundedness, and the validity of the Polyakov-Regge block decomposition. But it does preserve positivity so \eqref{cctildeinequality} still holds. Moreover, the flat-space limit $T^\text{trim}_\infty$ actually \emph{agrees} with $T^\text{sub}_\infty$ as long as we restrict ourselves to physical configurations $s > 4$ and $t,u < 0$. Indeed, for the part of this $s$-channel physical region that lies inside one of the $D_\mu$ we subtract exactly the right block to cancel the $E_\Delta^{(\ell)}$ divergences, whereas in its complement the contribution of these blocks vanishes in the flat-space limit. This also implies that, for physical $s$, the hyperbolic partial waves defined by both amplitudes agree; in the language of \eqref{hyppartialwavefromf} we may write
\begin{equation}
	f_\ell^\text{trim}(s) = f_\ell^\text{sub}(s)
\end{equation}
on their common domain of definition.

\subsection{Unitarity}
Now we make the following claim for the behavior of $T^\text{trim}$ inside the Mandelstam triangle:
\begin{equation}
\label{Tcutsubleading}
	\tilde{s}, \tilde{t}, \tilde{u}\geq0 \quad :\quad  \lim_{R \to \infty} \frac{\GG_c(\tilde{s},\tilde{\eta})}{|\GG_c( s,  \eta)|} T^{\text{trim}}(\tilde{s},\tilde{\eta}) = 0
\end{equation}
In the given domain there are divergences in $T^\text{trim}$ because (i) the cut-away $s$-channel conformal blocks also diverge for real $s < 4$ outside $D_\mu$, and (ii) in $T^\text{trim}$ the divergences in the images of $D_\mu$ under crossing have not been cancelled. These divergences are however easily verified to be offset by the ratio $\GG_c(\tilde s,\tilde\eta)/|\GG_c(s,\eta)|$, which can be simplified to the exponentially small term $(\tilde s/4)^{2 \Delta_\phi - 3/2}$.

Substituting equation \eqref{Tcutsubleading} into the definition of the reflected hyperbolic partial waves we see that only the disconnected part survives, which produces:
\begin{equation}
	\lim_{R \to \infty} \tilde c^\text{trim}_\ell(s) = i \sqrt{s} (s - 4)^{1-d/2}\,.
\end{equation}
and the inequality \eqref{cctildeinequality} can then be written as:
\begin{equation}
	1 \geq \left|1  + i \, s^{-1/2} (s- 4)^{d/2-1} f_\ell(s)\right|
\end{equation}
which is exactly the unitarity condition for flat-space partial waves $f_\ell(s)$.

\subsubsection{Asymptotic behavior}
The (hyperbolic) partial waves are not only bounded by the unitarity equation, but also by the existence of a dispersion relation at positive $u$. Indeed, we have shown that the amplitude $T^\text{sub}(s,u)$ is polynomially bounded; let us say it is less than $C s^N$ \footnote{Note that we have obtained polynomial boundedness without assuming anything about the distributional nature of the time-ordered correlation functions, as is necessary in the LSZ prescription}. But equation \eqref{positivedispersionrelation} then provides a bound for the OPE density at each spin $\ell$:
\begin{equation}
	\rho_\ell(s) <  C s^N / P_\ell^{(d)}(-1 + 2u / (4 - s))
\end{equation}
for any $0 \leq u \leq 2$. This is exactly the kind of falloff that immediately leads to the Froissart-Martin bound, as reviewed for example in \cite{tourdefrance}. (And, much as in the standard literature on the S-matrix, the exact statement of this bound are meant to be understood in an averaged sense rather than pointwise. We will discuss this in more detail in \cite{xiangpaper}.) Note that with more generous assumptions the same bound likely follows from a Mellin space analysis \cite{Haldar:2019prg}.

\section{Conclusions}

In the flat-space limit mild assumptions suffice to show that conformal four-point functions must reduce to scattering amplitudes that are consistent with unitarity and possess a large domain of analyticity. In upcoming work \cite{xiangpaper} we will discuss how the OPE density $c(\Delta,J)$ almost reduces to the partial waves $f_\ell(s)$, how the Lorentzian inversion formula can reduce to the Froissart-Gribov formula \footnote{The classical limits of the Lorentzian inversion formula have been discussed recently in \cite {Maxfield:2022nat}, for which $\D$ (and $\ell$) also become large.}, and the fate of the dispersive functionals of \cite{Caron-Huot:2020adz}. In the future it would be interesting, among many other things, to investigate the fate of correlators of unequal operators, with more than four points, or with a spectrum such that anomalous thresholds can occur. We would also like to gather evidence for the universality of our main assumption by numerically bounding the correlator in $E'$ (extending the results of \cite{Paulos:2021jxx}).

\section{Acknowledgments}
We would like to thank Aditya Hebbar and Shota Komatsu for collaborations at an early stage of this project. We are grateful to Jo$\tilde{\text{a}}$o Penedones for comments on the draft, and to Edoardo Lauria, Marten Reehorst, and Miguel F. Paulos for discussions. The authors are supported by Simons Foundation grant \#488649 (XZ) and \#488659 (BvR and XZ) for the Simons Collaboration on the non-perturbative bootstrap. XZ is also supported by the Swiss National Science Foundation through the project 200020 197160 and through the National Centre of Competence in Research SwissMAP.

\bibliography{biblio}

\clearpage
\onecolumngrid
\appendix
\setcounter{equation}{0}
\section{Supplemental Material: Conformal Mandelstam Variables}
\label{app: conformal Mandelstam}
In this appendix, we provide more details on the conformal Mandelstam invariants $(s,t,u)$, defined as \cite{Komatsu:2020sag}:
\begin{equation}
	r(s) = \frac{2 - \sqrt{s}}{2 + \sqrt{s}},\ \eta(s,t) = - 1 + \frac{2t}{4 -s}, \ u = 4 - t - s
\end{equation}
where $r = \sqrt{\rho \bar \rho}$, $\eta = (\rho + \bar \rho)/ (2 \sqrt{\rho \bar \rho})$ and $\rho$, $\bar \rho$ are the radial coordinates \cite{Hogervorst:2013sma} related to cross ratios $z,\bar{z}$ through
\begin{equation}
	\rho = \frac{1-\sqrt{1-z}}{1+\sqrt{1-z}},\quad \bar{\rho} = \frac{1-\sqrt{1-\bar{z}}}{1+\sqrt{1-\bar{z}}}\,.
\end{equation}

\begin{figure*}[!h]
\centering
\includegraphics[width=12cm]{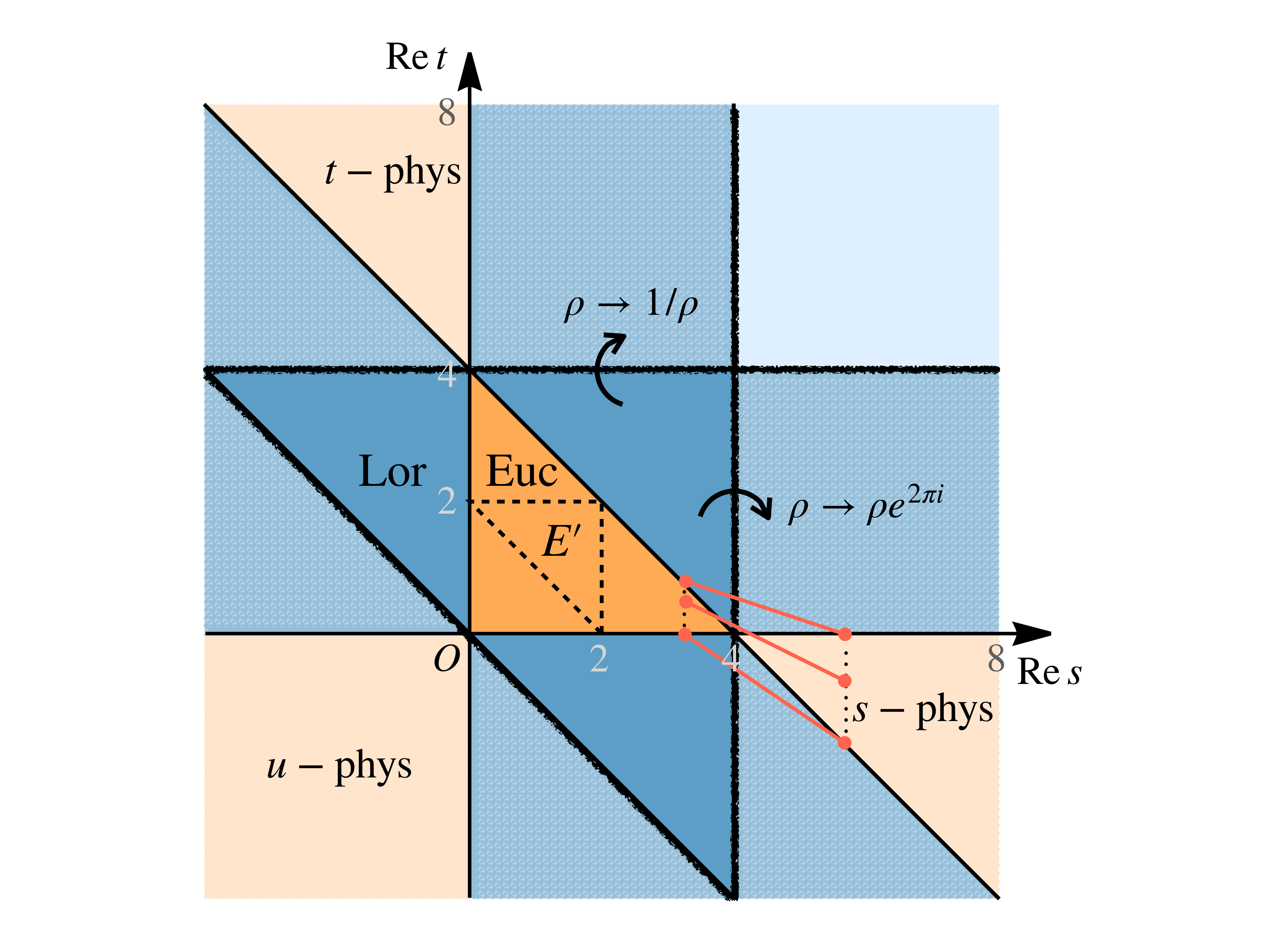}
\caption{The Mandelstam plane. The three triangles in the center with increasing size are $E'$, the small Mandelstam triangle (orange) and the large Mandelstam triangle (blue). Orange regions correspond to Euclidean regions where $\bar{z}=z^*\in\mathbb{C}$ up to an $e^{2\pi i}$ phase and the blue regions correspond to Lorentzian regions where $z,\bar{z}\in\mathbb{R}$ and independent. The fuzzy lines indicate branch cuts for the conformal correlator and the regions with lighter color correspond to secondary sheets, where all physical scattering configurations live. The top right light blue square corresponds to the `double discontinuity' region of the scattering amplitude, where no conformal block decomposition converges. The red lines indicate the involution map defined in \eqref{eq: involution} of the main text.}
\label{fig: Mandelstam plane}
\end{figure*}

The Euclidean configurations for the correlator ($z \in \mathbb C, \bar z = z^*$) are mapped to the small Mandelstam triangle $s,t,u \geq 0$ (see Figure \ref{fig: Mandelstam plane}). The spacelike separated Lorentzian configurations ($z,\bar z \in \mathbb{R}$ and independent) correspond to the remainder of the large Mandelstam triangle, \emph{i.e.} $s, t, u \leq 4$ and precisely one of $s, t, u$ non-positive. Physical scattering configurations can be found on secondary sheets, for example the points with $s > 4$ and $t,u < 0$ can be obtained from a Euclidean configuration by sending $\rho\to e^{2 \pi i}\rho$ while holding $\bar \rho$ fixed. The `double discontinuity' regions of the amplitude, \emph{i.e.} the region with $s, t > 4$ and its images under crossing, correspond to an analytic continuation where no conformal block decomposition converges. Note that these regions are not to be confused with the double discontinuity of conformal correlators defined below \eqref{confdispreln} in the main text.

\end{document}